# Automatic detection and diagnosis of sacroiliitis in CT scans as incidental findings

Yigal Shenkman[1], Bilal Qutteineh[2], Leo Joskowicz[1], Adi Szeskin[1],
Yusef Azraq[3], Arnaldo Mayer[4], Iris Eshed[5,6]

[1] The Rachel and Selim Benin School of Computer Science and Engineering, The Hebrew University of Jerusalem, Israel.

[2] Dept. of Orthopaedic Surgery, Hadassah Hebrew University Medical Center, Jerusalem, Israel.

[3] Department of Radiology, Hadassah Hebrew University Medical Center, Jerusalem, Israel.

[4] Computational Imaging Laboratory, Sheba Medical Center, Tel Hashomer, Israel.

[5] Department of Radiology, Sheba Medical Center, Tel Hashomer, Israel.

[6] Sackler School of Medicine, Tel Aviv University, Tel Aviv, Israel.

Corresponding author:
Prof. Leo Joskowicz
The Rachel and Selim Benin School of Computer Science and Engineering
The Hebrew University of Jerusalem
Edmond J. Safra Campus, Givat Ram, Jerusalem 9190401, Israel.
Email: josko@cs.huji.ac.il

# Abstract

Early diagnosis of sacroiliitis may lead to preventive treatment which can significantly improve the patient's quality of life in the long run. Oftentimes, a CT scan of the lower back or abdomen is acquired for suspected back pain. However, since the differences between a healthy and an inflamed sacroiliac joint in the early stages are subtle, the condition may be missed. We have developed a new automatic algorithm for the diagnosis and grading of sacroiliitis CT scans as incidental findings, for patients who underwent CT scanning as part of their lower back pain workout. The method is based on supervised machine and deep learning techniques. The input is a CT scan that includes the patient's pelvis. The output is a diagnosis for each sacroiliac joint. The algorithm consists of four steps: 1) computation of an initial region of interest (ROI) that includes the pelvic joints region using heuristics and a U-Net classifier; 2) refinement of the ROI to detect both sacroiliac joints using a four-tree random forest; 3) individual sacroiliitis grading of each sacroiliac joint in each CT slice with a custom slice CNN classifier, and; 4) sacroiliitis diagnosis and grading by combining the individual slice grades using a random forest. Experimental results on 484 sacroiliac joints yield a binary and a 3-class case classification accuracy of 91.9% and 86%, a sensitivity of 95% and 82%, and an Area-Under-the-Curve of 0.97 and 0.57, respectively. Automatic computer-based analysis of CT scans has the potential of being a useful method for the diagnosis and grading of sacroiliitis as an incidental finding.

# Introduction

Sacroiliitis is a condition resulting from inflammation of the sacroiliac joints (SIJ) (Vleeming et. al. 2012). It is frequently the first symptom of inflammatory diseases of the axial skeleton, e.g., ankylosing spondylitis, which often affects young patients, with mean age under 40 years (Calin et. al. 1994). It is characterized by lower back pain which can spread to the buttocks, legs, groin and feet. About 5% of patients with persistent low back pain have sacroiliitis (Calin et. al. 1994, Lawrence et. al. 1998). Early diagnosis of sacroiliitis can lead to preventive treatment which can significantly improve the patient's quality of life in the long run. However, when diagnosed late, the damage in the sacroiliac joint may be irreversible, and the patient may suffer from chronic lower back pain and limited mobility (Feldtkeller et. al. 2003).

The detection of sacroiliitis in its early stages is difficult and time-consuming, as the symptoms of sacroiliitis are similar to those of more common back conditions, e.g. a herniated intervertebral disc. As a result, the mean diagnostic delay of sacroiliitis is about seven years, and many patients remain undiagnosed (Lawrence et. al. 1998).

Patients with low back pain undergo physical examination to identify the source of their pain. When there is suspicion of sacroiliitis, an X-ray of the pelvic is acquired and evaluated by an expert musculoskeletal radiologist (Bennett et. al. 1968, Braun and Sieper 2007). However, it is known diagnosis based on X-rays has low sensitivity for detecting the early stages of the disease (van der Linen et. al. 1984, Braun and Sieper 2007, Montandon et. al. 2007).

Other imaging modalities are used by physicians for the diagnosis and grading of sacroiliitis. Bone scintigraphy is not recommended for sacroiliitis diagnosis (Zilber et. al. 2016), due to its low specificity (Yildiz et. al. 2001). CT is generally not recommended for the detection of sacroiliitis due to its relatively high radiation exposure. However, it may be used to identify sacroiliitis when the examination has already been acquired for other indications (Furtado et. al. 2005). MRI is the preferred diagnostic imaging modality for the detection of sacroiliitis due to its high contrast and tissue resolution (Puhakka et. al. 2003). However, MRI is expensive, time consuming and less available compared to the other imaging modalities.

The New York criteria, introduced in 1966 and revised in 1984, is the current standard for grading of sacroiliitis structural damage on X-rays (Bennett et. al. 1968, van der Linen et. al. 1984). It consists of five grades: 0 – *normal*, no disease; 1 – *suspicious*, some blurring of the sacroiliac joint margins; 2 – *mild sclerosis*, some erosions; 3 – *partial ankylosis*, severe erosions, reduced sacroiliac joint space; 4 – *complete ankylosis*, no sacroiliac joint space.

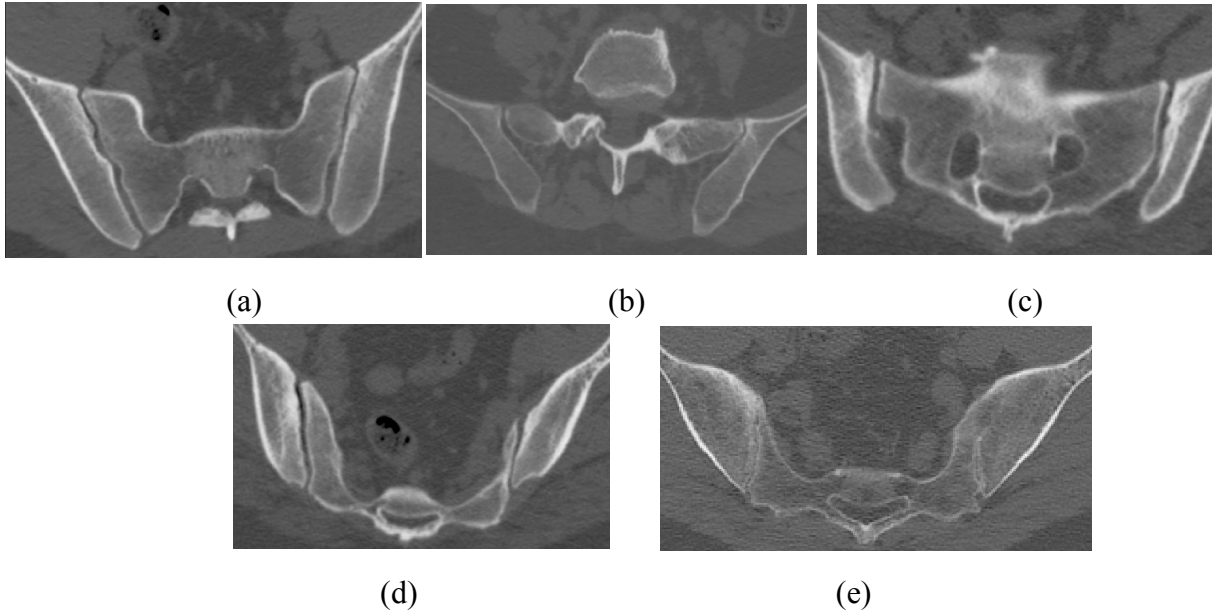

**Fig. 1:** Examples of axial CT slices showing sacroiliitis at different stages according to the New York criteria: (a) grade 0, normal, no disease; (b) grade 1, suspicious, some blurring of the joint margins; (c) grade 2, mild sclerosis, some erosions; (d) grade 3, erosions, manifested sclerosis, partial ankylosis; (e) grade 4, complete ankylosis, no sacroiliac joint space.

This grading criterion can be also used for grading CT scans (Fig. 1). However, Geijer et. al. 2009 indicates that the New York criteria might be unsuitable for sacroiliitis diagnosis in CT scans and proposes a simpler, more practical three-class grading: *no disease*, *suspected disease*, *definite disease*.

Sacroiliitis as a matter of incidental findings can be diagnosed in a CT scan when the scans have already been acquired, for other indications e.g. for lumbar spine examination or abdominal pain. However, it is not uncommon that the sacroiliac joints are overlooked when interpreting CT scans, leaving some patients undiagnosed. These patients may return for years later with worsened symptoms, but unfortunately, irreversible damage to the sacroiliac joint might already has occurred.

Incidental findings, defined as previously unknown, undiagnosed conditions that are unintentionally identified by clinicians during the evaluation of medical images, are oftentimes found in radiological examinations. Furtado et. al. 2005 report that 37% of patients with whole-body CT scan may have incidental findings. Thus, further evaluation of existing radiological images for various pathologies may lead to the early diagnosis of unknown conditions. However, this evaluation requires expertise, which is not always available on site, and increases the time required for the examination, which may not be feasible due to budget constraints.

In this paper, we address two key issues related to the diagnosis of sacroiliitis in CT scans: 1) the automatic detection and grading of sacroiliitis in CT scans as incidental findings, for patients who underwent lower back or abdomen CT scanning as part of their lower back workout, and; 2) an investigation of computer-based grading criteria of sacroiliitis in CT scans.

**Previous work**

To the best of our knowledge, there are no published methods for computer-based diagnosis (CAD) of sacroiliitis in CT scans. However, various methods have been developed for automatic detection and segmentation of structures and pathologies. Lately, there has been a shift towards deep learning-based methods, mostly due to their genericity and to the results they yield. Most of the recent methods focus on incidental findings because their identification does not require explicitly modeling the finding, only detecting that it is present in a scan. For a recent review of these methods, see Litjens et. al. 2017.

We review next methods for diagnosis and segmentation of musculoskeletal pathologies and structures, and then methods that specifically address computer-aided diagnosis of sacroiliitis.

Anthony et. al. 2016 describe a method for classifying knee osteoarthritis in X-rays into five grades. The method is based on transfer learning on a convolutional neural network (CNN) classifier pre-trained on ImageNet (Jia et. al. 2014). It achieves a 58% classification accuracy, an improvement upon the 29% achieved by *Wndchrm* (Shamir et. al. 2008) on 8,892 labeled X-ray images. Roth et. al. 2015 propose a method for sclerotic spine metastasis detection in CT scans. It consists of two steps: 1) lesion candidates' ROI generation with model-based methods, and; 2) candidate classification with a full CNN. They report a high false positive rate of ~50 patches per scan.

Chen et. al. 2015 introduce J-CNN, a CNN architecture to identify vertebrae in spine CT scans. The method combines ROI localization and vertebra identification in a single CNN with the sliding window technique. It achieves an identification rate of 84% on 302 labeled CT scans, an improvement upon the 74% rate reported by Glocker et. al. 2012 that uses regression forests and hidden Markov models. Shen et. al. 2015 also propose a method for localizing vertebrae with a CNN classifier with a sliding window. Their method achieves a mean 2.3 (std 1.6) pixel distance between computed vertebrae centers and the ground truth on 10 cases. An important drawback of the sliding window technique is that it incurs in high computation times.

Two studies address the specific task of computer-based detection and diagnosis of sacroiliitis. Huang et. al. 2010 describe a method for grading 2D scintigraphy images with a sacroiliac joint index (SII). It locates the pelvis by fuzzy subsets histogram thresholding, finds anatomical landmarks with model-based methods, refines the pelvis segmentation, and computes the SII by summing pixel values on horizontal lines. The sacroiliitis diagnosis is then performed manually based on the computed SII. However, Yildiz et. al. 2001 report that studies disagree about SII sensitivity for sacroiliitis. Chae et. al. 2017 describe a semi-automatic automatic deep learning method for diagnosing sacroiliitis on frontal X-rays. It relies on a manually defined Region of Interest (ROI) around the sacroiliac joint with which a CNN is trained. The method achieves a sensitivity of 62.3% and specificity of 57.2% on 3,362 X-ray radiographs.

In an earlier study, we made the first attempt to diagnose and grade sacroiliitis on CT scans (Hochman 2016). We developed an automatic method that combines model-based sacroiliac joint ROI localization and bag-of-words classification. It first segments the skeleton with the intensity threshold that creates the lowest number of connected components. It then identifies the pelvic region by searching for large in-plane (axial) changes in the convex hull of the segmented skeleton. The sacrum and the ilium are then segmented using the graph min-cut algorithm. Patches along the sacrum and ilium boundaries are then extracted and classified using a support vector machine (SVM). The sacroiliac joint is classified as *diseased* when the % of *diseased* patches exceeds a pre-set value. The method is unable to segment grade 4 cases; it achieves 62% classification accuracy on cases with successful segmentation (grades 0-3).

## Method

We present ***SIJ-grade***, a fully automatic algorithm for the diagnosis and grading of sacroiliitis in CT scans as incidental findings based on supervised machine learning and deep learning. The input is a lumbar CT scan. The output is a grade for the left and right sacroiliac joints (SIJ).

We first describe the classification labels, then present an overview of the method, and then detail each of the method steps.

**Classification labels**

We define three types of labels:

1. ***SIJ voxel label***: a binary label indicating if a voxel belongs to the sacroiliac joint or not.

2. ***SIJ slice grade***: indicates the grade (0-4) of sacroiliitis in an individual slice according to the New York criteria.
3. ***SIJ case grade***: indicates the grade (*healthy, suspicious, sick*) of the entire scan according to the criterion defined by Geijer et. al. 2009.

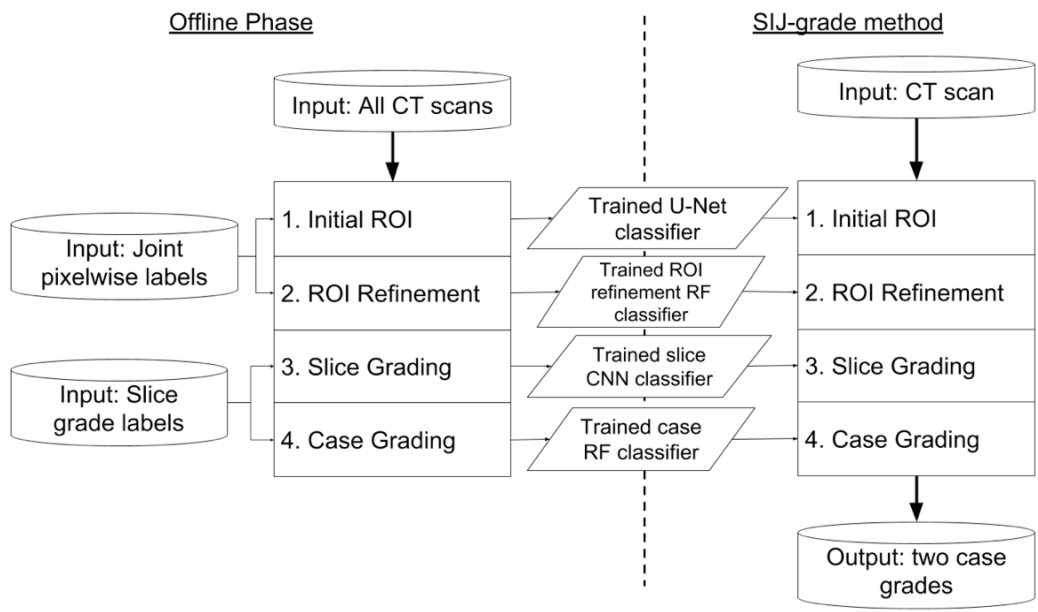

**Fig. 2:** *SIJ-grade* flow diagram of the offline training phase (left) and the online classification phase (right). The offline phase yields the U-Net classifier, the ROI refinement RF classifier, the custom slice CNN classifier, and the case RF classifier (center). *SIJ-grade* computes the grades of the left and right sacroiliac joints of a given CT scan. Both phases perform the same steps: 1) Initial ROI computation; 2) ROI refinement; 3) Slice grading; 4) Case grading.

These labels are used internally for the computation and allow us to explore the various sacroiliitis grading criteria, as described in the Results section (Data labeling).

**Method overview**

*SIJ-grade* uses supervised machine learning and deep learning techniques. It thus requires two phases: an ***offline training phase*** using labeled training and validation datasets, and an ***online classification phase*** for grading individual CT scans whose performance is evaluated with a labeled test dataset (Fig. 2). Both phases consist of four steps: 1) computation of an initial region of interest (ROI) that includes the pelvic joints region using heuristics and a U-Net classifier; 2) refinement of the ROI to detect both sacroiliac joints using a Random Forest (RF); 3) individual sacroiliitis grading of each sacroiliac joint in each CT slice with a custom slice CNN classifier, and; 4) sacroiliitis diagnosis and grading by combining the individual slice grades using a RF classifier. In the offline phase, each classifier is trained individually. In the online phase, the trained classifiers are used in tandem to produce the SIJ case grade.

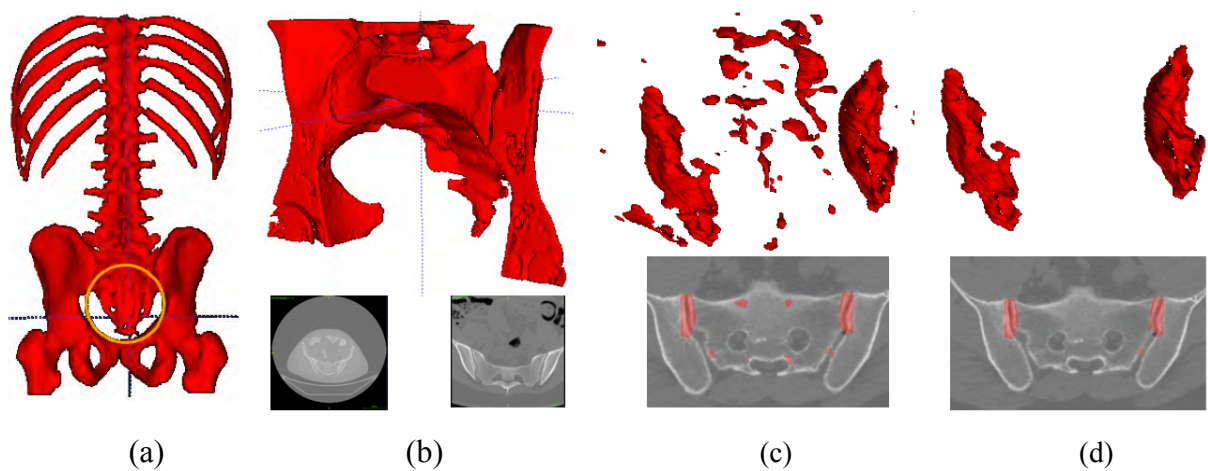

(a)　　　　　　　(b)　　　　　　　(c)　　　　　　　(d)

**Fig. 3:** Illustration of the ROI computation steps: (a) coarse skeleton segmentation (red) with coccyx region (yellow circle); (b) pelvis ROI (top, red) and two representative axial CT slices (bottom); (c) left and right sacroiliac joint ROIs and slice with initial ROI superimposed; (d) refined left and right sacroiliac joint ROIs and slice with its refined ROI (bottom, red).

**1. Initial ROI computation**

The initial ROIs of the left and right sacroiliac joints in each axial CT slice are computed in three steps: 1) coarse skeleton segmentation; 2) pelvis ROI computation, and; 3) left and right sacroiliac joint ROIs computation. Fig. 3 illustrates these steps, which we describe next.

The first step is the coarse segmentation of the skeleton structures by heuristic adaptive thresholding (Figs. 3a, b). It sets the upper threshold to 1,300 Houndsfield Units (HU) and finds the lower threshold by further thresholding the CT scan with 22 successive thresholds in the range of [150-500] HU; the threshold that yields the smallest number of connected components is chosen. Finally, a segmentation mask is computed with the upper and lower thresholds and with morphological closing along the three scan axes; a kernel diameter of 7 voxels was set experimentally.

The second step computes the pelvis ROI from the coarse skeleton segmentation (Fig. 3c). The top slice of the pelvis ROI is identified by computing the width of the convex hull of the skeleton segmentation for each axial slice starting at the bottom of the skeleton segmentation upwards. The top slice of the pelvis ROI is defined as the first slice in which the pelvis convex hull width is greater by 30% than that of the previous slice. The bottom slice of the pelvis ROI is determined by computing the pelvis convex hull in the slice above the top ROI pelvis slice, finding the first slice below it without bone voxels inside the pelvis convex hull, and adding a margin of 30mm (about 15 slices) below it.

The third step is the computation of the left and right sacroiliac joints ROIs (Fig. 3d) with a U-Net classifier (Ronnenberger et. al. 2015). The U-Net classifies the voxels in the pelvis ROI for each scan slice as SIJ voxels using three channels: the current slice, the slice before and the slice after it. The result is usually an over-segmentation of the sacroiliac joints.

**2. ROI refinement**

This step removes the falsely identified sacroiliac joint ROI voxels with an RF classifier. The inputs are the coarse skeleton segmentation, the initial ROI, and the CT scan. First, the coccyx is located; then, a volume consisting of the voxels inside the sacroiliac joint is computed with an RF classifier (Breiman 2001). The RF classifier classifies the voxels based on a features vector that includes the direction and magnitude of each voxel from the coccyx. It then intersects this volume with the voxels in the initial ROI and eliminates all connected component whose volume is < $2mm^3$ (value chosen empirically).

The coccyx location (Fig. 3a) is computed by evaluating two candidate locations: the backmost pelvic region segmentation voxels, and the backmost skeleton segmentation voxels, as the actual location depends on the scan region and on the quality of the pelvis ROI computation. The steps for coccyx localization are: 1) selection of the backmost pelvic region segmentation voxels as the first candidate coccyx location; 2) computation of a feature vectors for all voxels in this volume and classification with the RF classifier to obtain binary SIJ voxel labels; 3) intersection (logical AND) of the computed SIJ voxel labels with those computed by the U-Net classifier; 4) selection of the backmost skeleton segmentation voxels as the second candidate coccyx location; 5) steps 2-3 are repeated once to produce an additional intersection. The selected coccyx location is the one with the largest intersection computed in steps 3 and 5.

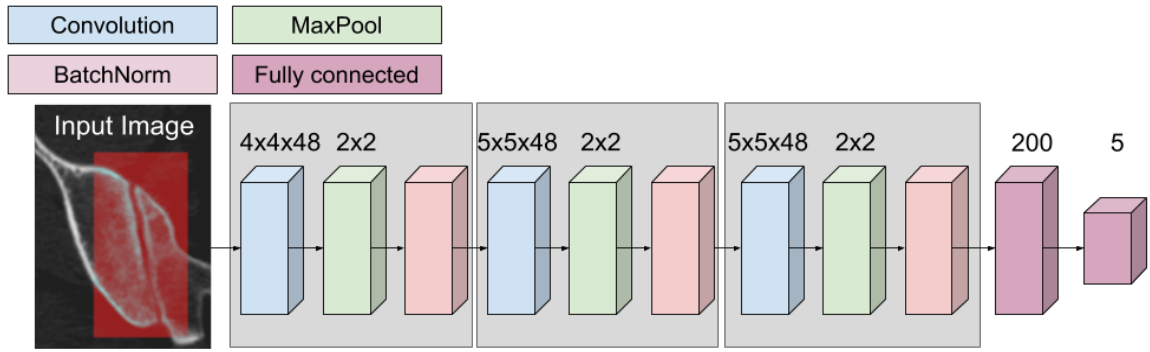

**Fig. 4.** Architecture of the custom slice CNN classifier. The input is a 100 x 200 pixels axis-aligned rectangular ROI image (red). The network consists of three blocks (gray boxes) and two fully-connected layers. Blue boxes represent convolutional layers feature maps with ReLU activation (the maps sizes are shown above them). Green boxes represent max pool layers (the filter sizes are shown above them). Red blocks represent batch normalization layers. Purple boxes represent fully connected layers (their sizes are shown above them). The first fully connected layer has ReLU activation, and the last fully connected layer has softmax activation.

**3. Slice grading**

Individual slices that include the left and right sacroiliac joints are assigned an SIJ slice grade with a custom slice CNN classifier (Fig. 4). In the preprocessing step, an axis parallel rectangle is computed for each joint based on the refined ROI by vertically splitting the axial slice images into two half-slice images, each containing one sacroiliac joint. The left side half-slice image is then flipped horizontally and an axis-parallel rectangle centered at the refined ROI center is computed from each half-slice image.

The custom slice CNN consists of 11 layers organized in three blocks of a convolution feature map with a ReLU activation layer, a max pool layer and a batch normalization layer followed by two fully-connected layers. The first fully connected layer performs ReLU activation, and the last fully connected layer perform softmax activation. The final layer's size is five, to match the five SIJ slice grades. All layers' activations are the ReLU activations except for the last fully-connected layer which performs softmax activation.

## 4. Case grading

In the final step, an SIJ case grade is computed for each sacroiliac joint based on the five-class SIJ slice grades in the ROI according to a new empirical criterion developed by Drs. B. Qutteineh and Y. Azraq:

*Slice-to-case SIJ grading* **criterion**

a. *sick*: more than one slice has an *SIJ slice grade* 4 **OR** at least three consecutive slices have an *SIJ slice grade* 3.
b. *suspicious*: the *SIJ case grade* is not *sick* **AND** at least a third (~30%) of the slices have a *SIJ slice grade* 2.
c. *healthy*: the *SIJ case grade* is neither *sick* nor *suspicious* according to criteria a. and b.

The SIJ case grade is automatically computed with a RF classifier based on a feature vector computed from a SIJ slice grade vector. The SIJ slice grade vector is defined by the consecutive SIJ slice grades in the ROI; its length is the number of slices in the ROI. The feature vector consists of the length of the longest consecutive sequence in the SIJ slice grade vector for each SIJ case grade. For example, for the SIJ slice grade vector $sgv$ = 01123333322310 (SIJ grades of 14 consecutive slices in the SIJ ROI), the entries in the feature vector for SIJ slice grade 0 is 11 (0 at the beginning and 0 at the end of $sgv$), for SIJ slice grade 1 is 21 (two 1 s at the beginning and one at the end of $sgv$), for SIJ slice grade 2 is 21 (two 2 s at the end and one at the beginning of $sgv$), for SIJ slice grade 3 is 51, and for SIJ slice grade 4 is 00 (no 4 slices appear in $sgv$). Note that we use an RF approach based on the "argmax" of each SIJ case grade instead of the Slice-to-case SIJ grading criterion to accommodate for possible slice mis-labelings and discrepancies.

**Data augmentation for offline training**

We performed the following data augmentations for the offline training phase of the classifiers on the training set for each one of the four steps of the method.

<u>ROI computation</u>: for the initial ROI U-Net training, each axial CT slice was augmented by: 1) random scaling in the range of 1±0.15; 2) random rotation in the range of ±2º; 3) random width translation in the range of ±10% of the slice's width; 4) random height translation in the range of ±1% of the slice's height; 5) elastic deformation by small image distortions (Simard et. al. 2017). The distortions are generated with a random displacement field by first randomly moving each pixel along each axis uniformly in the range of ±1 pixel, smoothing the displacement field with a Gaussian with a standard deviation $\sigma = 10$, and scaling the resulting displacement field by a factor of 5. No random flip was performed since the half-slices have already been flipped. No data augmentation was used for the ROI refinement RF classifier.

Slice grading: for the custom slice CNN classifier, the individual axial slices were augmented by: 1) random scaling in the range of 1±0.15; 2) random rotation in the range of ±10º; 3) random width translation in the range of ±10% of the slice's width; 4) random height translation in the range of 0-1% of the slice's height.

Case grading: for the case RF classifier, new feature vectors were generated for each SIJ case in the training set by augmenting each slice in the case ROI with the slice grading augmentation described above with the same parameter settings for all slices and classifying it with the custom slice CNN classifier. The feature vector for each such augmentation was then computed as described in the Case grading subsection above. In total, 20 SIJ slice augmentations were performed, yielding 20 SIJ new slice grade vectors with which the RF classifier was trained.

## Experimental results

We conducted a comprehensive set of experiments with three goals: 1) evaluate the performance of the *SIJ-grade* method with various settings; 2) quantify the accuracy of the computer-based grading criteria; 3) quantify the intra-observer case grading variability.

### Datasets

We obtained 242 anonymized axial CT scans from the Sheba Medical Center, Ramat Gan, Israel (Prof. I. Eshed). The total number of slices is 50,941, with a mean of 210.5 (range 152-675) slices per CT scan. Axial plane slices have 512×512 voxels of sizes in the range of 0.25×0.25×0.33mm$^3$ to 3×3×3mm$^3$. A third of the scans (81) include the spine and pelvis only; two-thirds (161) include the body from the neck to the knees. All scans included a case grade assigned by Prof. I. Eshed adopted from the five-class New York grading criterion (Bennett et. al. 1968, van der Linen et. al. 1984).

### Datasets labeling

We defined three types of labels: 1) *SIJ voxel* label; 2) *SIJ slice grade*, and; 3) *SIJ case grade*. The SIJ voxels were manually labeled slice by slice with ITK-SNAP by Y. Shenkman. The *SIJ voxel* labels were then validated in 12 scans (8 scans of patients with partial/complete ankylosis and 5 scans randomly selected) by Dr. B. Qutteineh. The *SIJ slice grades* for each sacroiliac joint were assigned with a custom GUI following axial CT slice inspection by Dr. Qutteineh. The *SIJ case grades* were automatically computed from the *SIJ slice grades* according to the criterion previously described in sub-section 4 (Case grading).

|  | *SIJ voxel* | | *SIJ slice grades* | | | | | *SIJ case grades* | | |
|---|---|---|---|---|---|---|---|---|---|---|
| *SIJ case grades* | 0 | 1 | 0 | 1 | 2 | 3 | 4 | healthy | suspicious | sick |
| **Study grades** | sacroiliac joint ROI | | *slice two-classes* | | | | | *SIJ two-classes* | | |
|  | | | 1 | 0 | 4 | 3 | 2 | healthy | suspicious | sick |
|  | | | *slice three-classes* | | | | | *SIJ three-classes* | | |
|  | | | 1 | 0 | 2 | 4 | 3 | healthy | suspicious | sick |
|  | | | *slice five-classes* | | | | | Two observers variability study | | |
|  | | | 0 | 1 | 2 | 3 | 4 | | | |

**Table 1**: Grading labels: ***SIJ voxel*** (0,1), ***SIJ slice grade*** (0-4) and ***SIJ case grade*** (*healthy*, *suspicious*, *sick*). ***SIJ case grades*** are computed for three slice classes and two case classes. The row under each class shows the classes partition according to the ***SIJ case grades***.

Each axial slice was split into two half slices, yielding a total of 34,894 half slices. The grade distribution is: 11,552 grade 0, 12,574 grade 1, 5,302 grade 2, 5,209 grade 3, 257 grade 4. From a total of 484 cases, the SIJ case grades distribution is: 198 *healthy*, 58 *suspicious*, 228 *sick*.

We define three classification groups of the SIJ slice grades (Table 1):
1. *Slice two-classes*: class 0 includes SIJ slice grades 0 and 1, class 1 includes *SIJ slice grades* 2, 3, 4.
2. *Slice three-classes*: class 0 includes SIJ slice grades 0 and 1, class 1 includes *SIJ slice grade* 2, class 2 includes *SIJ slice grades* 3 and 4.
3. *Slice five-classes*: One class for each *SIJ slice grade* 0-4.

We define two classification groups of the SIJ case grades:
1. *SIJ two-classes*: class 0 is *healthy*, class 1 is *unhealthy* (*suspicious* or *sick*).
2. *SIJ three-classes*: class 0 is *healthy*, class 1 is *suspicious*, class 2 is *sick*.

The *SIJ voxel* labels are used to define the ground truth ROI for each sacroiliac joint in each axial slice and to train the initial ROI and refinement ROI U-Net and RF classifiers. The ROI rectangle size was defined to be 50x25mm$^2$. The *SIJ slice* grades are used to train the custom slice CNN classifier. The *SIJ case* grades are used to train the case RF classifier.

**Observer variability**

To establish a baseline for observer variability, we compared the five-class case grades originally assigned by Prof. I. Eshed (IE) to the SIJ case grades obtained from the manual slice five-classes grades assigned by Dr. B. Qutteineh (BQ). The SIJ *two-* and *three-class* grades were computed using the *slice-to-case grading* criterion described above. The agreement between both observers for the *SIJ two-*and *three class* grades is 77.4% and 65.7%, respectively. For the *SIJ two-class grades*, the distribution is: *healthy* 186/279, *unhealthy* 260/167 for BQ and IE respectively. For the *SIJ three-class grades*, the distribution is: *healthy* 186/279, *suspicious* 47/70, *sick* 213/97 for BQ and IE respectively. Note that Dr. B. Qutteineh *SIJ case grades* are on average higher than those of Prof. I. Eshed.

**Training, validation and test sets partition**

The labeled datasets were partitioned into disjoint training, validation, and test sets for each of the method steps as follows.

ROI localization: the training set for the U-Net classifier for the initial ROI step consists of 50 cases (100 SIJs, 2,579 slices). The validation set consists of 8 cases (16 SIJs, 652 slices). The training set for the RF classifier for ROI refinement step consists of 13 cases (26 SIJs, 3,720 slices) randomly selected from the U-Net classifier's training set. The validation set is used to evaluate the performance of each one of the trained classifiers and of the *SIJ-grade* method.

Slice and case grade classification: the training set for the U-Net, refinement RF, slice CNN and case RF classifiers consists of 181 cases (362 SIJs, 12,621 slices). The validation set consists of 30 cases (60 SIJs, 2,667 slices). The test set consists of 31 cases (62 SIJs, 2,158 slices). Data augmentation for all the training sets was performed as described in Section 2. The validation set is used to evaluate the performance of each one of the trained classifiers. The test set is used to evaluate the performance of the *SIJ-grade* method.

Note that the exact partition of the data into training, validation, and test set is not very important and will most likely yield very similar results. What is important is that the sets are mutually disjoint, that they are chosen randomly, and what are the approximate sizes of each set. The training required training sets with fewer tagged examples than are usually required in machine learning and deep learning methods: 20% (100 out of 484) for ROI localization, 5% (26 out of 3,720) for ROI refinement. For slice and case grading, it is 75%. A similar observation applies for validation sets: 3% for ROI localization (16 out of 484) and 8% (2,667 slides out of 34,894) for slice grading and 12% (60 out of 484) for case grading.

**Evaluation methodology**

We evaluate the performance of the offline training and the ***SIJ-grade*** method with five studies.

1. ROI localization accuracy. For the offline training phase, both steps of the ROI localization were evaluated. For the ***SIJ-grade*** method, the computed axis-aligned ROI of each sacroiliac joint was compared to the ground-truth ROI defined by the manually labeled SIJ voxels with the Dice coefficient between the two ROI rectangles and the distance between their centers.

2. Slice and case grading accuracy. For the offline training phase, the custom slice CNN classifier and the case RF classifier were trained as follows: 1) train the custom slice CNN classifier for one epoch; 2) compute a SIJ slice grade vector for each sacroiliac join in the training and validation sets by using the ROI rectangles computed by the custom slice CNN classifier; 3) compute a feature vector from the SIJ slice grade vector; 4) train the case RF classifier on the feature vectors and compute its accuracy on the validation set, and; 5) repeat steps 1-4 until the custom slice CNN classifier training converges. The resulting RF classifier consists of 500 trees, each of depth four.

The custom slice CNN and the case RF classifiers were tested in six different scenarios. A custom slice CNN classifier is trained for each of the three *Slice two-, three-,* and *five-classes* groups. For each of these custom slice CNN classifiers, a case RF classifier is trained for each of the two *SIJ two-* and *three-classes* classification groups.

For the ***SIJ-grade*** method, a case RF classifier was trained for each epoch of the custom slice CNN classifier. The case RF classifier with the highest validation accuracy was then selected to evaluate the test set. The SIJ grades accuracy is tabulated with confusion matrices.

3. Case grade classification robustness. For the offline training phase, two alternative case RF classifier were tested: the *latent space* CNN classifier with full feature vectors and the *one-hot* CNN classifier with simplified feature vectors. The input to the CNN classifiers is the output of the last fully connected layer of the slice CNN classifier (Fig. 4). It is an embedding of an ROI rectangle into a latent space whose dimension $m$ is the number of SIJ slice grade classes (2,3,5). It is computed for all SIJs in the training set and accumulated into a $k \times m$ matrix where $k$ is the number of slices in the ROI. For the *latent space* classifier, the rows of the matrix are the latent space embeddings of each slice. For the *one-hot* classifier, the row entries of the latent space matrix are binarized by setting the maximal cell value to 1 and the rest to 0.

For the **SIJ-grade** method, we compared the accuracy of the case RF classifier to that of six alternative classifiers: *SIJ two- and three-classes* case classification with *slice two-, three*, and *five-classes* classification. We select the best 10 case RF classifiers and rank them by their accuracy on the validation dataset and compute their accuracy on the test set.

4. <u>Cross validation</u>. To evaluate the robustness of the **SIJ-grade** method to various training, validation and test dataset splits, we split the entire dataset of 484 SIJs with the same balance: 362 SIJs (75%) for training, 60 SIJs (12%) for validation, and 62 (13%) SIJs for testing. We performed 6-fold cross validation on these sets. We evaluated the slice and case grading by randomly splitting the training and validation cases into six sets and computed the accuracy of each of the six RF case classifiers on the test set for each of the classes.

We define an *ensemble case classifier* as follows. In each cross validation fold, six case RF classifiers are trained, one for each data split. Each classifier output is interpreted as a grade probability vector whose size is the number of case classes; the value of the cell $i$ in this vector is the probability that the case grade is the class. The *ensemble case classifier* computes six prediction vectors and sums them. The *SIJ case grade* is the one of the cell with the maximum value. The classifier's accuracy and confusion matrices are then computed on the test set.

| | *SIJ two-classes* grading accuracy (%) | | *SIJ three-classes* grading accuracy (%) | |
|---|---|---|---|---|
| **Slice grades** | case RF classifier | slice CNN classifier | case RF classifier | slice CNN classifier |
| *slice two-classes* | 82.3 | 81.8 | 77.4 | 79.7 |
| *slice three-classes* | 87.1 | 79.0 | 80.6 | 79.2 |
| *slice five-classes* | 89.3 | 64.2 | 71.7 | 61.6 |

**Table 2**: Accuracy of the case RF classifiers and their corresponding slice CNN classifiers. Rows are for slice grading slice CNN classifiers with *slice two-, three-* and *five-classes*. Columns show the test accuracy of case RF classifiers for *SIJ two-* and *three-classes* case classification, with the corresponding slice grading classifier test accuracy.

| **Computed *SIJ three-class* grade** | | | | | | | | | |
|---|---|---|---|---|---|---|---|---|---|
| **True *SIJ three-class* grade** | *slice two-classes* | | | *slice three-classes* | | | *slice five-classes* | | |
| | healthy | suspicious | sick | healthy | suspicious | sick | healthy | suspicious | sick |
| *healthy* | **0.80** | 0.04 | 0.16 | **0.92** | 0 | 0.08 | **0.92** | 0 | 0.08 |
| *suspicious* | 0.14 | **0.29** | 0.57 | 0.29 | **0.29** | 0.42 | 0.43 | **0** | 0.57 |
| *sick* | 0.13 | 0 | **0.87** | 0.17 | 0 | **0.83** | 0.27 | 0.03 | **0.70** |

**Table 3**: Confusion matrices of *SIJ-grade* for the *SIJ two-class* based on three slice classes.

| **Computed *SIJ two-class* grade** | | | | | | |
|---|---|---|---|---|---|---|
| **True *SIJ two-class* grade** | *slice two-classes* | | *slice three-classes* | | *slice five-classes* | |
| | healthy | unhealthy | healthy | unhealthy | healthy | unhealthy |
| *healthy* | **0.92** | 0.08 | **0.88** | 0.12 | **0.80** | 0.20 |
| *unhealthy* | 0.24 | **0.76** | 0.14 | **0.86** | 0.14 | **0.86** |

**Table 4**: Confusion matrices of *SIJ-grade* for the *SIJ three-class* based on three slice classes.

5. <u>Case sensitivity-specificity trade-off</u>. To evaluate the sensitivity-specificity trade-off of the ***SIJ-grade*** method, we evaluated two scenarios: *threshold tuning* and *two-step classification*. For *threshold tuning*, the case RF classifier's output is the probability Pr(*grade*) of belonging to the corresponding class. The *SIJ two-classes* grading is tuned with a predefined threshold $\tau \in [0,1]$: when the case RF classifier determines that Pr(*unhealthy*), the classifier predicts *unhealthy*, *healthy* otherwise. Similarly, the *SIJ three-class* grading is tuned with two thresholds, $\alpha \in [0,1]$ and $\beta \in [-1,1]$ such that when Pr(*suspicious*) > $\alpha$, the grade is *suspicious;* else, when Pr(*sick*)–Pr(*healthy*) > $\beta$, the grade is *sick;* else, the grade is *healthy*. For *two-step classification*, the cases graded as *healthy* are identified first, and the remaining ones are re-classified as *suspicious* or *sick*.

|  | *SIJ two-classes* grading accuracy (%) | | | *SIJ three-classes* grading accuracy (%) | | |
|---|---|---|---|---|---|---|
| **Slice grades** | latent-space | one-hot | slice CNN classifier | latent-space | one-hot | slice CNN classifier |
| *slice two-classes* | 71.0 | **82.3** | 81.8 | 66.1 | 69.4 | **79.7** |
| *slice three-classes* | 80.6 | 79.0 | **79.0** | 66.1 | 66.1 | **79.2** |
| *slice five-classes* | **79.0** | 77.4 | 64.2 | 53.2 | 53.2 | **61.6** |

**Table 5**: Test accuracy of the *latent space* and *one-hot* classifiers. Rows are for slice grading slice CNN classifiers with *slice two-, three-* and *five-classes*. Columns show the test accuracy of the *latent space* and *one-hot* classifiers for *SIJ two-* and *three-classes* case classification, with the corresponding slice grading classifier test accuracy.

**Results**

1. <u>ROI localization accuracy</u>. The **SIJ-grade** method achieves on the validation dataset a mean Dice coefficient of 0.82 and a mean ROI centers distance of 4.9mm (std 4.7mm).

2. <u>Slice and case grading accuracy</u>. Table 2 shows the accuracy of the **SIJ-grade** method for the best case grading classifier on the test set. It achieves accuracies of 87.1% and 80.6% accuracy for *SIJ two-* and *three-classes* grading.

Table 3 shows the confusion matrix for *SIJ two-classes grading*. The **SIJ-grade** method misses 24% of the *unhealthy* cases in the *slice two-classes* classification and 14% in the *slice three-* and *five-classes* classification. Table 4 shows the confusion matrix for *SIJ three-classes* grading. The **SIJ-grade** method misclassifies 71% of the suspicious cases in the *slice two-* and *three-classes* classification, and all the *suspicious* cases in *the slice five-classes* classification.

| Slice grades | SIJ two-classes grading accuracy (%) | | | SIJ three-classes grading accuracy (%) | | |
|---|---|---|---|---|---|---|
| | *latent-space* | *one-hot* | *case RF* | *latent-space* | *one-hot* | *case RF* |
| *slice two-classes* | 77(4) | 76(4) | **81(4)** | 60(11) | 71(3) | **73(3)** |
| *slice three-classes* | 81(3) | 77(6) | **84(5)** | 59(7) | 66(3) | **74(4)** |
| *slice five-classes* | 80(4) | 77(5) | **82(3)** | 64(8) | 64(8) | **67(4)** |

**Table 6:** Mean (standard deviation) test set accuracy of the 10 best case RF classifiers and their corresponding latent-space, one-hot and slice CNN classifiers. Columns are as in Table 5.

| | SIJ two-classes grading accuracy (%) | | | SIJ three-classes grading accuracy (%) | | |
|---|---|---|---|---|---|---|
| *slice two-classes* | 82.3 | 82.3 | 87.1 | 67.7 | 74.2 | 80.6 |
| | 82.3 | 77.4 | 91.9 | 72.6 | 66.1 | 74.2 |
| | **std=4.6** | | | **std=4.7** | | |
| *slice three-classes* | 83.9 | 75.8 | 90.3 | 67.7 | 67.7 | 79.0 |
| | 88.7 | 80.6 | 83.9 | 69.4 | 69.4 | 74.2 |
| | **std=4.8** | | | **std=4.1** | | |
| *slice five-classes* | 80.6 | 67.7 | 75.8 | 69.4 | 67.7 | 74.2 |
| | 75.8 | 77.4 | 82.3 | 72.6 | 66.1 | 72.6 |
| | **std=4.6** | | | **std=2.9** | | |

**Table 7:** Mean and std accuracies of the 6-fold cross validation on the test set. Rows show slice grading slice CNN classifiers with *slice two-, three-* and *five-classes*. Columns show the classification accuracy of the *latent space* and *one-hot* classifiers for *SIJ two-* and *three-classes*.

| Case grades | SIJ two-classes grading accuracy (%) | SIJ three-classes grading accuracy (%) |
|---|---|---|
| *slice two-classes* | 90.3 | 80.6 |
| *slice three-classes* | 91.9 | 77.4 |
| *slice five-classes* | 80.6 | 71.0 |

**Table 8**: Accuracy of the *ensemble case classifier*. Rows show case grading classifiers with *slice two-, three-* and *five-classes*. Columns show the test accuracy of case RF classifiers for *SIJ two-* and *three-classes* case classification.

|  | **Computed *SIJ two-class* grade** | | | | | |
|---|---|---|---|---|---|---|
| **True *SIJ two-class* grade** | *slice two-classes* | | *slice three-classes* | | *slice five-classes* | |
|  | *healthy* | *unhealthy* | *healthy* | *unhealthy* | *healthy* | *unhealthy* |
| *healthy* | **0.92** | 0.08 | **0.88** | 0.12 | **0.72** | 0.28 |
| *unhealthy* | 0.11 | **0.89** | 0.05 | **0.95** | 0.14 | **0.86** |

**Table 9**: Confusion matrices of the *ensemble case classifier* for *SIJ two-class* grades for the *two-, three* and *five-slice classes*.

|  | **Computed *SIJ two-class* grade** | | | | | | | | |
|---|---|---|---|---|---|---|---|---|---|
| **True *SIJ three-class* grade** | *slice two-classes* | | | *slice three-classes* | | | *slice five-classes* | | |
|  | *healthy* | *suspicious* | *sick* | *healthy* | *suspicious* | *sick* | *healthy* | *suspicious* | *sick* |
| *healthy* | **1.00** | 0 | 0 | **0.96** | 0 | 0.04 | **0.84** | 0 | 0.16 |
| *suspicious* | 0.43 | **0** | 0.57 | 0.29 | **0** | 0.71 | 0.29 | **0** | 0.72 |
| *sick* | 0.17 | 0 | **0.83** | 0.13 | 0 | **0.87** | 0.2 | 0.03 | **0.77** |

**Table 10**: Confusion matrices of *ensemble case classifier* for *SIJ three-class* grades for the *two three* and *five-slice classes*.

3. <u>Case grade classification robustness.</u> Table 5 shows the accuracies of the *latent-space* and *one-hot case* grading classifiers. The case RF classifier achieves a mean 80.5% accuracy, 11.2% higher than the *latent-space* classifier that achieves 69.3% mean accuracy, and 6.9% higher than the *one-hot* classifier that achieves 73.7% mean accuracy.

Table 6 shows the accuracies of the case RF classifier to the best 10 case RF classifiers for six alternative classifiers. The case RF classifier outperforms the *latent-space* and the *one-hot* classifiers (mean and std) in all six scenarios, which indicates that the RF classifier is robust.

4. <u>Cross-validation.</u> Table 7 shows the test dataset accuracies and standard deviations for each fold in the 6-fold cross validation. The low (< 5%) standard deviations indicate that the classifiers are robust and do not depend on the dataset split.

Table 8 shows the accuracy of the *ensemble case classifier*. The classifier achieves the highest accuracy with *SIJ two-classes* classification, 91.9%, an improvement of 4.2% over the case RF classifier that achieved 87.1%, but no improvement for *SIJ three-classes* classification in which both the case RF classifier and the *ensemble case classifier* achieved 80.6% accuracy.

Tables 9 and 10 show the confusion matrices of the *ensemble case classifier*. The sensitivity improved from 86% for the case RF classifier by 9% to 95% for the *ensemble case classifier* for SIJ two-classes classification, but for *SIJ three-classes* classification the sensitivity for class *suspicious* declined from 29% to 0%.

5. <u>Case sensitivity-specificity trade-off</u>. Fig. 5 shows the ROCs (Receiver Operating Characteristic) curves of the SIJ two-class grade classification computed from True Positive and False Positive rates for different values of $\tau$ (Mossman 1999). Note the curves show a high TP rate close to 1 and a low FP rate close to 0. This is not the case the SIJ three-class grade classification that is computed from the True Positive (TP) and False Positive (FP) rates of each grade for different values of $(\alpha, \beta)$ for which there is no clear trend for various parameter value combinations.

Table 13(a) shows the confusion matrix for the representative threshold value $\tau = 0.42$ with an accuracy of 88.7%. Table 13(b) shows the confusion matrices for the representative threshold values $(\alpha, \beta)$ of the *ensemble case classifier* for the *threshold tuning* and *two-step* classifications for $(\alpha, \beta) = (0.14, 0)$.

6. <u>Accuracy with respect to observer variability</u>. The **SIJ-grade** method accuracy with respect to either observer grading is 94.8% for the *SIJ two-class* grade (up from 91.9% for Dr. Qutteineh) and 87.9% (up from 80.6% for Dr. Qutteineh) for the *SIJ three-class* grade.

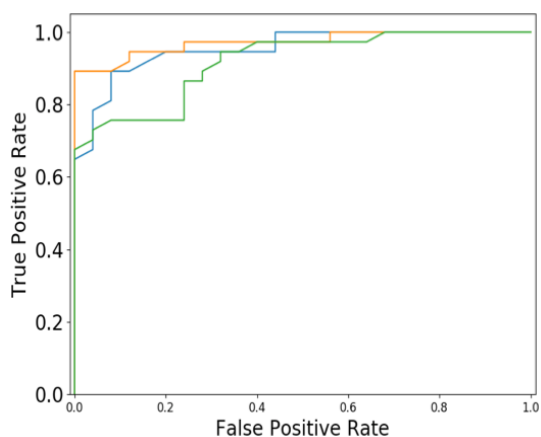

**Fig. 5**: Two-way ROC curve of the *SIJ two-classes* grade classification computed from True Positive and False Positive rates for different values of $\tau$. The Area-Under-the-Curve (AUC) is 0.95 for the *slice two-classes* (blue curve), 0.97 for the *slice three-classes* (orange curve) and 0.82 for the *slice five-classes* (green curve) grades.

**Implementation and running time**

The offline training and **SIJ-grade** classification methods were implemented as a pipeline of four modules. This allows for executing and changing modules individually, and for running parallel multi-threaded pipelines. The code is written in MATLAB and Python and runs both on Linux and Windows. It uses the Python libraries *dicom2nifti*, *NiBabel* (NIPY developers 2016), *NumPy* (Oliphant 2015), *SciPy* (Oliphant 2007), *PyTables* (PyTables developers 2002), *scikit-image* (van der Walt 2014), *TensorFlow* and *Keras* (Abadi et. al. 2015) to train the U-net and slice CNN classifiers and *scikit-learn* (PyTables developers 2002) to train the ROI refinement and case RF classifiers (Pedregosa et. al. 2011). The implementation uses GPUs and Nvidia's CUDA library to speed up training (Nickolls et. al. 2008).

The running time of **SIJ-grade** on a single CT scan is about 45 seconds on an Intel i7-5930K 3.50GHz CPU, 32GB RAM, Nvidia TITAN Xp GPU. The offline phase took 15-16 hours for the *slice two-classes* and *slice three-classes* classifiers and 28-29 hours for the *slice five-classes* classifiers for 181 cases with 12,621 half-slices.

| $\tau = 0.42$ | Computed *SIJ two-class* grade (88.7%) | |
|---|---|---|
| **True *SIJ two-class* grade** | *slice two-classes* | |
| | healthy | unhealthy |
| healthy | **0.76** | 0.24 |
| unhealthy | 0.03 | **0.97** |

| $(\alpha, \beta) = (0.14, 0)$ | Computed *SIJ three-class* grade threshold (72.6%) | | |
|---|---|---|---|
| **True *SIJ three-class* grade** | *slice three-classes* | | |
| | healthy | suspicious | sick |
| healthy | **0.64** | 0.36 | 0 |
| suspicious | 0 | **0.86** | 0.14 |
| sick | 0 | 0.23 | **0.77** |

| | Computed *SIJ three-class* grade two-steps (80.6%) | | |
|---|---|---|---|
| | *slice three-classes* | | |
| | healthy | suspicious | sick |
| | **0.92** | 0.36 | 0 |
| | 0 | **0.86** | 0.14 |
| | 0 | 0.23 | **0.77** |

(a)  (b)

**Table 13**: Confusion matrices: (a) *SIJ two-class* based on *two-slice classes* with τ =0.42 (88.7% accuracy); (b) *SIJ three-class* based on *slice three-classes* with thresholds $(\alpha,\beta)=(0.14,0)$ (center, 72.6% accuracy) and two-steps (right, 80.6% accuracy).

## Discussion

To the best of our knowledge, **SIJ-grade** is the first fully automatic method for the detection and grading of sacroiliitis in CT scans as incidental findings, for patients who underwent lower back or abdomen CT scanning as part of their lower back workout. The method is based on a hybrid approach consisting of machine learning methods and deep learning methods for high-quality sacroiliac joint ROI identification, individual slice classification, and case classification. Our experimental setup and results lead to the following observations.

The *SIJ case* grades agreement between both observers of 77.4% for the *slice two-class* and 65.7% for the *SIJ three class* grades indicate a significant intra-observer variability. Two possible reasons are: 1) the difference between the specialty and expertise of the observers -- one observer is a senior musculoskeletal radiologist (IE) and the second is a junior orthopaedic surgeon (BQ), and; 2) the junior observer graded individual slices, while the senior observer graded the entire case. The case grades of the junior observer were computed with *Slice-to-case SIJ grading* criterion, who also confirmed that the case grading was correct. Direct slice grading comparison was infeasible due to the senior observer time limitations. We conclude that the binary *SIJ-case* grading (*healthy/unhealthy*) is best suited for the classification, especially for incidental findings warning.

The ROI localization accuracy (Dice score of 0.82 and ROI center distance of 4.9mm with std of 4.4mm) is significantly better than that of our model-based method (Dice score of 0.53 and ROI center distance of 21.3mm with std 15.5mm) (Hochman 2016). This indicates that an accurate ROI localization is important to achieve a high SIJ case grading score accuracy.

The **SIJ-grade** method accuracy for the *SIJ two-classes* (91.9%) and *SIJ three-classes* (80.6%) is well above the inter-observer variability (77.4% and 65.7%, respectively) and model-based method accuracy: 61.7% for *SIJ two-classes* grading, 74.9% for image patches of sacroiliac joints with grade 3, and 55-60% for the remaining grades. The confusion matrices indicate that the *unhealthy* grade cases for *SIJ two-class* grade have a low misclassification rate: 5% for the *slice three-classes,* 11% for the *slice three-classes* and 14% for the *slice five-classes*. In contrast, the *suspicious* grade cases for *SIJ three-class* grade have an unacceptable misclassification rate: 71% for the *SIJ two-* and *three-class* and 100% for the *SIJ five-classes*. The two-step classifier misclassification rate is 8,14 and 30% for grades *healthy, suspicious,* and *sick* respectively. This indicates that the **SIJ-grade** method is better at binary classification.

The slices grade classification robustness results indicate that the *slice CNN* classifier outperforms both the *latent-space* and *one-hot* classifiers in all categories for the *SIJ three classes*, and is very close to them for the *SIJ two classes* except for the *slice five-class* case. This shows that a satisfactory discrimination power was achieved and that the training set examples were sufficient for slice-based classification. The accuracy and std of the case RF classifier in six scenarios for both *SIJ two-* and *three-class* grades is very similar, which indicates that it is robust (Table 7). Similarly, the 6-fold cross validation results show a low standard deviation, which indicates that the classifier is robust to various datasets partitions.

The *ensemble case classifier* version of **SIJ-grade** outperforms all other variations. It achieves 91.9% accuracy, the highest of *SIJ two-classes* classification and an improvement of 4.2% over the case RF classifier. However, it showed no improvement for *SIJ three-classes* classification (80.6% accuracy). Note that both are well above the intra-observer variability.

The sensitivity-specificity tradeoff experiments indicate that the **SIJ-grade** method achieves a very high AUC of 0.95 and 0.57 for the *slice two-* and *slice-three* classes. The confusion matrices indicate a good discrimination power for all cases for *SIJ two-* and *three-class* grades.

Our experimental results also shed light on the issue of CT sacroiliitis grading. Following (van der Linen et. al. 1984,) we chose to grade cases into two or three classes, and not five. However, individual slice grading was done for five classes to allow for finer discrimination; case grading was then performed by grouping into 2 and 3 case classes. The *slice-to-case SIJ case grading* criterion follows what was observed in how radiologists detect and grade sacroiliitis a CT scan. It consists of the examination slice by slice of each sacroiliac joint. When there is a clear indication of deterioration -- one or more slices show complete ankylosis with no sacral joint space (grade 4), or three consecutive slices show partial ankylosis and severe erosions, partial joint space (grade 3), there is definite sacroiliitis, with grade 3 or 4. If none of these apply, then when at least a third of the slices show mild sclerosis and/or some erosions -- then there is mild sacroiliitis (grade 2). Otherwise, it is healthy, since the grades of 0 and 1 are suspicious, which is hard to differentiate. This distinction into binary or three-way grading is confirmed by the experimental results, which show that *slice five-classes* underperform the other groupings.

The **SIJ-grade** method accuracy with respect to either observer grading is 94.8% for the *SIJ two-class* grade**,** up from 91.9% for both observers, and 87.9%, up from 80.6% for both observers for the *SIJ three-class* grade. This indicates that the **SIJ-grade** method has high accuracy despite the observer variability and that it is not fitted to a single observer.

While the computational methodology and experimental results presented in this paper address sacroiliitis detection and grading, we believe that the issues, methodology, experiments design and results provide insights on how to approach radiological disease grading issues in similar clinical problems. Indeed, radiological grading of scans into a few, empirically defined grades is ubiquitous. Often times, for volumetric scans, e.g. CT and MRI, the case grading is obtained from the slice grading with an empirical, unstated heuristic. The grading that is used for the individual slices may differ from that of the cases, as it is the case in our research on sacroiliitis. Our experiments explore this issue and help to quantitatively determine what are the case grades that yield the best quantitative results. Moreover, the proposed pipeline is typical for this type of grading problems: finding a Region of Interest (ROI), grading individual slices, and then grading the entire case. From an experimental point of view, the use of deep learning methods for ROI identification and individual slices classification, the augmentations performed, and the number of labeled training examples that are necessary to achieve the desired accuracy provide insights and indications for what to expect in similar grading problems.

## Conclusion

We have developed the first fully automatic method that detects and grades sacroiliitis on CT scans as incidental findings for CT scans as incidental findings, for patients who underwent lower back or abdomen CT scanning as part of their lower back workout. Our method is intended to be used to alert radiologists about sacroiliitis in a CT scan regardless of the reason for its acquisition. It grades individual axial slices in CT scans and implements a new criterion for sacroiliitis case grading based on individual slice grading. The individual slice grading enables high case grading accuracy with relatively few data sets and achieves significantly better results than our model-based method.

**Acknowledgments:** This research was supported in part by Grant 53681 from the Israel Ministry of Science, Technology and Space and by a Hebrew University Grant on Personalized Medicine. We thank Eliel Hochman and Clara Herscu for providing part of the heuristic Region of Interest code and Prof. M. Liebergall, head of the Dept. of Orthopaedic Surgery, Hadassah Hebrew University Medical Center for facilitating this research.

**Conflict of Interest**: none of the authors has any conflict of interest. The authors have no personal financial or institutional interest in any of the materials, software or devices described in this article.

**Protection of human and animal rights statement**: no animals or humans were involved in this research. All scans were anonymized before delivery to the researchers.

**Presentation at conferences**: Abstract of a preliminary version of this paper presented at the Int. Symp. on Computer Aided Radiology and Surgery, 2017 and at the MICCAI 2018.

**Conformance with publisher sharing policy**: This manuscript is published under the CC-BY-NC-ND license. The final paper will be published in Medical Image Analysis, vol. 57, in October 2019. DOI link: https://doi.org/10.1016/j.media.2019.07.007